\documentclass[twocolumn]{aa} %
\usepackage{graphicx,color}
\usepackage{txfonts}
\begin{document}
   \title{Models for Weak Wind and Momentum Problems in the Winds of Hot Stars}

  \author{O. Vilhu
         \inst{1}
          \and
          T.R. Kallman
     \inst{2}
          }

  \offprints{O. Vilhu}

   \institute{Dept of Physics, P.O. Box 84, FI-00014  University of Helsinki, Finland\\
              \email{osmi.vilhu@gmail.com}
         \and
                 Lab. for High-Energy Astrophysics, NASA/GSPC, Greenbelt, Maryland, USA
       \email{timothy.r.kallman@nasa.gov}
                 }

   \date{Received ; accepted  }

  \abstract
{Hot star winds  are laboratories for 3-dimensional radiative hydrodynamics and for X-ray sources with wind accretion. In this context analytic models presented here are helpful.}
{ 
 The  CAK-method (Castor, Abbot  \& Klein, 1975)  has been succesful for
 giants and supergiants of normal OB-stars but has failed to explain the weak winds of main 
sequence low luminosity OB-stars (‘weak wind problem’). Further, CAK has never been 
applied seriously to WR-stars and was considered as a mission impossible due to the
 ‘momentum problem’.     
The aim is to reevaluate the analytic CAK-method, to  recalculate proper force multipliers, numerically solve the wind equations for a sample of O- and Wolf Rayet (WR) -stars  and to obtain their mass loss rates and wind velocities . The secondary aim is to solve the $\it{weak \, wind \,  and \, momentum \,  problems}$ of hot star winds.}
{The wind in the supersonic part was  modelled by photoionized plasma and radiative force (force multiplier) using the XSTAR-code  (Kallman \& Bautista, 2001, Stevens \& Kallman, 1990). The force multiplier $FM$ was computed as a function of the absorption parameter $t$ , ionizing parameter $\xi$, particle number density $N$, chemical composition and the ionizing source spectrum. The force was included in the momentum equation, and together with the mass conservation solved numerically in the supersonic part of the wind for a sample of O- and  WR- stars (WN-type).  The input parameters were the basic stellar parameters (mass, radius, luminosity, chemical composition). The results depend also on the boundary condition of subsonic part and the velocity law. Fitting  with the   $\beta$-law with fixed $\beta$=0.6 and v$_{in}$ = 10 km/s approximate these and  gave the mass loss rate and wind velocity as outputs.  Mass  clumping was introduced by the volume filling factor $F$$_{vol}$  scaling $\xi$  by  $F$$_{vol}$$^{-1}$. Velocity clumping was approximated by the velocity filling factor $FVEL$ modifying the force multiplier  (following Sundqvist et al., 2014). }
{Force multipliers based on blackbody radiators can be used for O-stars, while cut blackbodies (flux below 230 Å cut to zero) approximate well those of WR-stars. O-stars require moderate clumping  $F$$_{vol}$ = 0.13 to match the  canonical Vink-prediction (Vink et al. 2001). The low mass loss rates of main sequence late O-stars (weak wind problem) can be explained by velocity clumping ($FVEL$ = 0.1). The   momentum problem of WR-stars is shown to be due to wrong treatment of the input ionizing spectrum  resulting in  too small force multiplier. Due to heavy absorption in WR-winds the flux below 230 Å  (He II ionization) is zero enhancing  greatly the number of absorbing  heavy element lines, and consequently the force multiplier, by eliminating the suppression   by soft X-rays. The computed mass loss rates and terminal wind velocities for 40 OB-stars and 55 WR-stars (WN type) are given in Tables A1 and A2 and Figs. 10-14.
     }
{A possible solution for the weak wind problem of low luminosity late O-stars was quantitatively studied  and explained by a small velocity filling factor $FVEL$. The momentum problem of WR-winds was solved by proper computation of the  line force with correct radiator (cut to zero below 230 Å). The problem is   an opacity problem of simply identifying enough lines  (Gayley et al. (1995)).  The present paper is  the first  comprehensive and self consistent treatment and numerical solutions of hot star wind equations. Starting from the basic stellar parameters (mass, radius, luminosity, chemical composition)  the wind equations  were solved fitting with the $\beta$-law (with fixed $\beta$=0.6) giving mass loss rate and wind velocity as  results.  The computed mass loss rates match well with the observed/predicted ones. The effects of free parameters $\beta$, $F$$_{vol}$ and  $F$$_{vel}$=$FVEL$ were quantitatively estimated. The eventual  X-ray suppression on the face-on side of Cyg X-3 may act like lowering of $FVEL$, leading to decreasing of both mass loss rate and wind velocity.  }


   \maketitle
%

\section{Introduction}

Hot star winds are accelerated by the radiation pressure in lines.  The properties of such winds were first comprehensively explored by Castor, Abbott \& Klein,  1975, hereafter CAK.   This theory is based on the Sobolev approximation to compute the local line force, that is, that the line broadening is dominated by the bulk motion of the wind, and that photons from the stellar photosphere only interact once as they escape the wind. 

 CAK showed that the effects of line scattering can be represented by the ratio of the opacity produced by the ensemble of lines to the electron scattering opacity and can be represented by a number, called the force multiplier, and that this quantity can be very large ($\geq 10^4$) in hot stars.  The formalism developed by CAK has been succesful in describing the wind properties, i.e. wind speeds and mass loss rates, in  giants and supergiants of normal OB-stars.  However, it does not provide correct predictions for the mass loss rates of main sequence low luminosity OB-stars (hereafter called the ‘weak wind problem’). Furthermore, the CAK formalism has not been applied to Wolf-Rayet (WR) stars, owing to the fact that winds from these stars appear to have more momentum than is available from the stellar radiation field (hereafter called the  ‘momentum problem’). 
Wind models which do not rely on the Sobolev approximation have been developed in order to produce  2-D simulations (see e.g. Sundqvist et al. 2018). In these simulations density and velocity clumpings arise in a physical way.  

Explanations which have been suggested for the 'weak wind problem' include the effects of X-rays (Marcolino et al. 2009), magnetic fields (Shenar et al. 2017)  or velocity clumping (Sundqvist et al. 2014). Leakage of light associated with porosity in velocity space could lead to lowering of starlight power (line force)  and consequently to lower mass loss rate. This last possibility will be studied in detail in the present paper. 

Another problem with the CAK-theory has been its inability to explain the massive  winds of Wolf Rayet stars.  This is called  the  'momentum problem'. 
The name comes from the ratio of wind momentum to radiation momentum $\eta$ =$M$$_{dot}$$v$$_{inf}$/($L$/c) which is much  larger than 1 for WR-stars. However, as pointed out by Gayley et al. (1995), the problem may be just due to the problem of identifying enough lines.   If this is the case then the CAK force multiplier generally accepted is too small.  In the present paper the force multipliers  are recalculated using realistic radiation spectra for WR-stars  and found that  this is the case (at least for the WN subclass of hottest WR stars). 

In the present paper we use the extensive XSTAR code  and data base  (Kallman \& Bautista, 2000; Stevens \& Kallman, 1990) to compute force multipliers.  The wind equations (mass conservation and momentum balance) for a sample of O- and WR-stars will be  solved in one dimension.  Inputs include basic stellar parameters: mass, radius, luminosity and chemical composition.  The wind equations are integrated, and the resulting velocity law is fitted with  an analytic form which is similar to the formula derived by CAK, and which is in widespread use for describing hot star winds.  The output solutions are the two quantities describing the wind global properties: mass loss rate and terminal velocity.  We also use the analytic  modification of the line force developed by Sundqvist et al. (2014) to include velocity clumping in the formalism.  Density clumping is included by using a volume filling factor modifying the ionization parameter  $\xi$.  

We discuss the derived mass loss rates and terminal velocities  in the context of other estimates and examine the effects of clumping in density and velocity.

\section{ The Line Force (Force multiplier)
 }
\subsection{Method of computation}

The force multiplier $FM$ was computed following the method described by CAK. An important difference is that CAK (and other compilations, eg. Gayley et al. 1995) assume that the ionization and excitation in the wind is given by a Saha-Boltzmann distribution modified by an analytic dilution factor applied to all elements.  We compute the ionization distribution in the wind by balancing the ionization and recombination due to the stellar radiation field.  We also employ the Boltzmann distribution to determine the populations of excited levels.    The radiation field is computed using a single-stream integration outward from the stellar photosphere.  In this way we calculate the line force appropriate to any spectral form of the photospheric spectrum and at any layer in the wind. This permits us to integrate wind equations from the surface to infinity  in a self-consistent manner since the upwards force is known.
  
The ionization balance and outward transfer of photospheric radiation are calculated using the  XSTAR photoionization code  (Stevens \& Kallman, 1990; Kallman \& Bautista, 2000; Kallman, 2018).  The ion fractions at each spatial position in the wind are used to calculate the force multiplier $FM$ by summing the CAK line force expression over an ensemble of lines.  The list of lines is  taken from Kurucz (http://kurucz.harvard.edu/linelists.html) and permits us to use local wind parameters. The line list is more extensive than in the original CAK-work.  It is assumed that the wind is spherically symmetric around the donor star (the ionizing source). 

XSTAR  calculates the ionization balance for all the elements with atomic number Z $\leq$ 30 together with  the radiative equilibrium temperature. 
It calculates full non-LTE level populations for all the ions of these elements.  It includes a fairly complete treatement of the level structure of each ion, i.e. more than $\sim$50 levels per ion, and up to several hundred levels for some ions.  Many relevant processes affecting level populations are included, i.e. radiative decays, electron impact excitation and ionization, photoionization, and Auger decays.  All processes include their inverses such that they obey detailed balance relations and populations approach LTE under the appropriate circumstances. The electron kinetic temperature is calculated by imposing a balance between heating from fast photoelectrons and Compton scattering with radiative cooling.

 Key simplifications employed by XSTAR are with regard to the radiative transfer solution.  Escape of line radiation is treated using an escape probability formalism, and this affects the temperature via the net radiative cooling.  Transfer of the ionizing continuum is treated using a single-stream integration of the equation of transfer, including opacities and emissivities calculated from the local level populations etc.  Thus there is no allowance for the inward propagation of diffusely emitted radiation.  This last approximation is likely to be most imortant for the results presented here, since strong winds can have large optical depths in the ionizing continua of hydrogen and He II.  It would be desirable to verify our current results using a transfer solution which does not have this limitation.   Limitations of the escape probability assumption for line escape have been pointed out by eg. Hubeny (2001).  However, for hot star winds the strong velocity gradient reduces most line optical depths to $\sim$a few at most, and so the lines are effectively thin and the radiative cooling is not strongely affected.

 The force multiplier $FM(t, \xi, N)$ is  a 3-dimensional function of  the absorption parameter $t$, ionizng parameter $\xi$ and particle number density $N$:
\begin{equation}
  t = \sigma_{\mathrm{e}}v_{\mathrm{th}}\rho(dv/dr)^{-1} 
\end{equation}
\begin{equation}
    \xi = \frac{L_{\mathrm{ly}}}{Nr^{2}}\ .
\end{equation} 

Here  $L_{ly}$ is the  incident luminosity below  the Hydrogen ionization limit 912 Å,  $\sigma_{e}$ is the elctron scattering coefficient, $\rho$ and $N$ the gas   and particle number density, respectively, $v$ is the outward wind velocity, $r$ the radial distance from the ionizing source and $v_{th}$ is the gas thermal velocity (typically $\sim$10 km/s). The force multiplier was computed  assuming a point source radiator.  When applied to stars,  the dilution factor $r^{-2}$ in eq.2 should be replaced by  the finite disk one (eq.15).
The chemical abundances of the wind, as well as the ionizing source spectrum,  can be specified as input data.  The method is in principle  the same as in the classical work of CAK. However,  in the present work $\xi$ and $t$ are explicitly included and computed locally in the wind. 

The force multiplier depends on $t$ and on the cross section for line absorption:
\begin{equation}
$$FM(t,\xi)=\Sigma_{lines}{\frac{\Delta\nu_DF_\nu}{F} \frac{1}{t}(1-e^{\eta t})}$$
\end{equation}

where $\Delta\nu_D$ is the thermal Doppler width of the line, $F$ is the total flux in the continuum, $F_\nu$ is the monochromatic flux at the line energy, and 
\begin{equation}
$$\eta=\kappa_{line}/\sigma_e$$
\end{equation}
\begin{equation}
$$\kappa_{line}=\frac{\pi e^2}{m_ec} g_L f \frac{N_L/g_L-N_U/g_U}{\rho\Delta\nu_D}$$
\end{equation}
where $N_L$ and $N_U$ are the lower and upper level populations, $g_i$  are the statistical weights, and $f$ is the oscillator strength.  Our code (xstarfmult.f) uses the ion fractions calculated as described above to sum over lines and calculate $FM(t, \xi)$.  Ground state level populations are taken directly from XSTAR;  excited level populations  are calculated  assuming a Boltzmann (LTE) distribution.  This is necessitated by the  use of the  Kurucz line list, which does not include collisional rates or other level-specific quantities which would allow a full non-LTE calculation of populations which can be used for the calculation of $\kappa_{line}$.  Most previous calculations of the CAK force multipliers have employed the LTE assumption for excited levels as well (eg. CAK, Abbott 1982,  Gayley 1995).

An extensive set of 
of $FM(t, \xi,N)$-grids were computed with  varying chemical composition, ionizing source spectrum and   particle number density $N$.  Dependence on the gas density is primarily through the ionization parameter $\xi$.  In our computations gas thermal velocity of Hydrogen atoms was used. As noted by CAK, Abbott (1982) and Stevens \& Kallman (1990), the force multiplier is independent of thermal velocity if the same value is used for both $FM$-tables and  $t$-parameter calculations. 

\begin{figure}
   \centering
   \includegraphics[width=9cm]{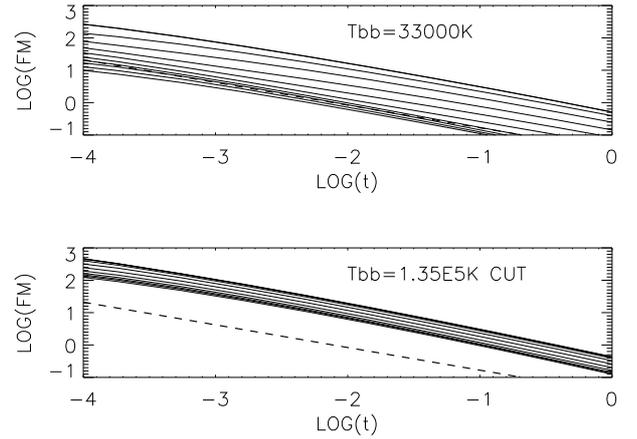}
      \caption{ Examples of  force multipliers.  Solar abundances were used and the ionizing source is a blackbody with T =  33 000 K (O-stars, upper plot) and  a cut blackbody with T=135 000 K (WR-stars, lower plot, see Fig.3). The logarithm of the ionizing parameter $\xi$ range  from 0 (uppermost curve) to  4.5 (lowest).    The  CAK-formula $FM$ = $1/30\times t^{-0.8}$ is shown by the dashed line. A density value of  $N=10^{12} cm^{-3}$ was used in this calculation. }
         \label{fm1}
   \end{figure}

\begin{figure}
   \centering
   \includegraphics[width=9cm]{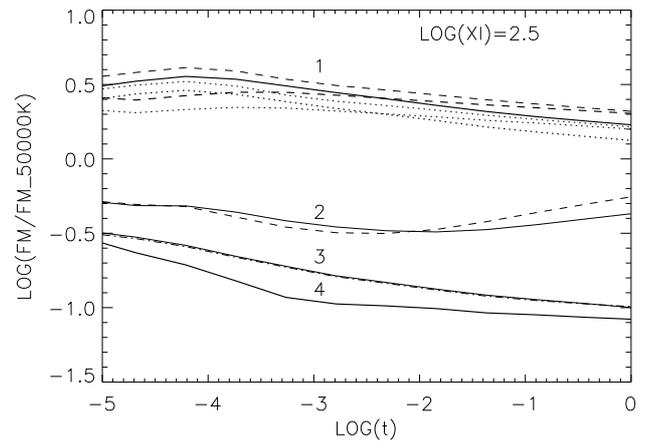}
      \caption{ Force Multipliers relative to 50 000 K blackbody at log($\xi$) = 2.5 using solar abundances.   The curves are grouped and labeled  with the ionizing source type as follows (see  Fig.\ref{sources}) : 1: blackbody with  T = 135 000 K but  cut to zero  below 230 Å  ( solid line), Potsdam model WNE16-20 (dashed line) and the mean  computed WR-model (the present study) at $\delta$(r/R$_{star}$) = 0.01 above the surface (dashed line).
 2: blackbody with  T= 33 000 K (solid line) and Potsdam model OB33-36 (dashed line).
3: blackbody with  T = 100 000 K (solid line)  and the same blackbody but with WN-abundances (dashed line, hardly visible). 
 4:  blackbody with  T = 135 000 K .  The dotted lines show force multipliers for the Set  1 with Hydrogen deficient abundances (see the text). }
         \label{fm2}
   \end{figure}

\begin{figure}
   \centering
   \includegraphics[width=9cm]{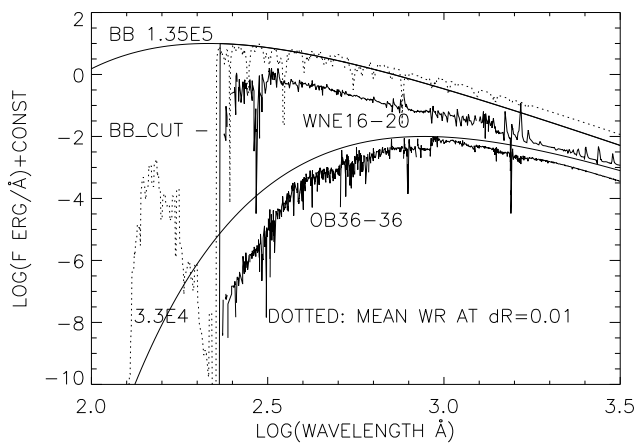}
      \caption{The ionizing source spectra used for force multipliers in  Fig.\ref{fm2}.  blackbodies  with T = 135 000 K (cut and non-cut) and T = 33 000 K are shown by continuous solid lines. The Potsdam WR-model WNE16-20   and OB-model OB33-36  show strong absorptions (solid lines). The dotted line is the mean spectrum of  WR-stars  at 1 percent (in stellar radius) above the surface (the present study). The curves can be shifted vertically. }
         \label{sources}
   \end{figure}

Examples of the force multiplier are shown in Fig.\ref{fm1} computed with solar abundances (Asplund et al. 2009) and $N$=10$^{12}$ cm$^{-3}$.  Comparison with the  widely used CAK-formula   $FM$ = $1/30 \times t^{-0.8}$ shows agreement at high ionization parameter values for 33 000K blackbody.  The CAK calculations used a smaller list of lines, which is the likely explanation for the fact that the dashed line generally lies below the current calculations. the Effects of other parameters are illustrated in  Fig.\ref{fm2}.  The ionizing source spectra used in Figs 1 and 2 are shown in  Fig.\ref{sources}.

\subsection{Effect of chemical abundances}

The difference between solar abundances  (by mass  X (Hydrogen)= 0.71, Y(Helium) = 0.27, Z(Heavy elements) = 0.015) and Hydrogen deficient WN-abundances (X = 0.085, Y = 0.9 , Z = 0.015)  is  moderately small if for individual heavy elements the same solar abundances  are used. This is demonstrated in Fig.2 for the Set 1 of radiators, the dotted lines showing force multipliers for Hydrogen deficient abundances. The effect of  equilibrium CNO-abundances in WN stars is negligible.   Fig.\ref{fm2} was computed with solar abundances  but the dashed line in the set 3 shows the line force for equilibrium CNO-abundances (hardly visible).  The effect is small, only at very low $t$-values (log($t$) less than -5) the effect is meaningful. 

 At small  $t$-values the force multiplier is proportional to Z, at large  $t$-values the Z-dependence is weaker.  
  
\subsection{Effect of  radiator spectral shape}

\begin{figure}
   \centering
   \includegraphics[width=9cm]{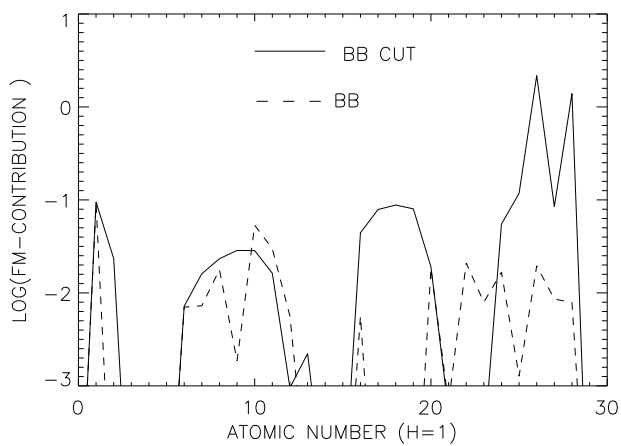}
     \caption{Contributions to the force multiplier ($FM$) as a function of the element atomic number (26 for Fe). 135 000 K blackbody (dashed line) and cut blackbody with the same temperature (solid line)  were used as  radiators and the force multiplier was calculated at log($\xi$) = 2.5 and log($t$) = -1.84 (see Fig. 2).  Note the logarithmic scale of y-axis.  }
         \label{elemcontri}
   \end{figure}

\begin{figure}
   \centering
   \includegraphics[width=9cm]{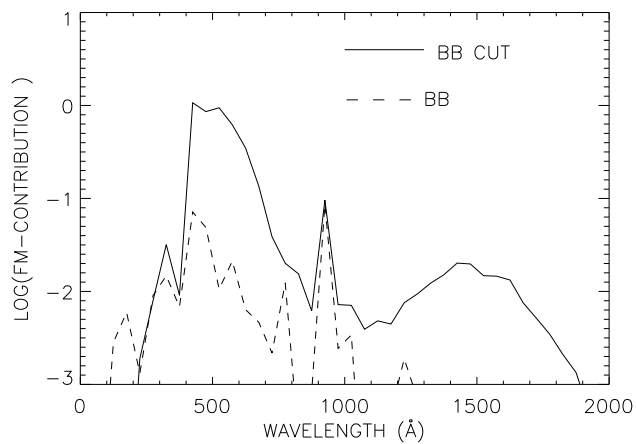}
     \caption{The same as in Fig. 4 but contributions to the force multiplier  inside  50 Å wide bands are shown as a function of wavelength.  }
         \label{wlcontri}
   \end{figure}

The spectral shape  below 912 Å, particularly below the Helium 2nd ionization limit 230 Å, is crucial. This can be seen in  Fig.\ref{fm2} by comparing the curves drawn by solid lines and labeled by '4' (blackbody T = 135 000 K) and  by '1'  (the same blackbody T = 135 000 K but cut to zero below 230 Å).  The cut part represents soft X-rays which heavily suppress the force multiplier as  found already by Stevens \& Kallmann (1990).   Fig.\ref{fm2} shows quantitatively how the force multiplier of  cut 135 000 K blackbody is over 10 times larger than that of a pure blackbody. Figures  \ref{elemcontri} and  \ref{wlcontri} show this effect as a function of element atomic number and wavelength. The soft X-ray suppression clearly vanishes when flux below 230 Ä is absorbed away, making the iron group elements contribution larger.

 The effect of blackbody temperature is moderately large.  Force multipliers of blackbodies between 25 000 K - 50 000 K  do not differ from those using corresponding Kurucz-models (Kurucz, 1979) for an average O-star.  However, force multipliers computed with 50 000 K and 25 000 K Kurucz models are somewhat smaller and larger, respectively,  than those with  corresponding blackbodies.
 
When integrating the equations of motion   the  absorption was included. This  changed the spectral form and luminosity below 912Å, the limit below which the ionization parameter $\xi$  is computed.  WR-stars with massive winds develope rapidly, close to the surface,  a spectral form where  fluxes at  short wavelengths (in particular below 230 Å, HeII ionization edge) are significantly  reduced.  Already at a few percent (of stellar radius) above the surface, the flux below 230 Å drops by a factor over  10.  At the same time the overall  reduction below 912Å is just by a factor of 2 or less. This is due to the large increase in absorption below 230 Å.   For O-stars, with  smaller mass loss rates, this effect is negligible. Hence, in Fig.2 the 33 000 K blackbody and the Potsdam model of the same temperature OB33-36 (Hainich et al. \citeyear{hainich3}, T = T$_{star}$ = 33 000 K) are very similar.  
 
The mean WR-spectrum  at $\delta$(r/R$_{star}$) = 0.01  and the force multiplier based on it  are included in Figures   \ref{fm2} and  \ref{sources}.
 The result is almost the same for the Potsdam model  WNE16-20 (Todt et al.  \citeyear{todt}, T = 141 300 K). Hence, the force multiplier based on cut blackbody represents well the force multiplier based on realistic WR-spectrum (see Ch. 3.3). 

The momentum/opacity problem is illustrated in Figs, 4 and 5 using 135 000K blackbody as the radiator (both pure BB and cut BB, spectrum cut to zero below 230 Å). The cut blackbody gives over 10 times larger contribution to the force multiplier than the pure BB, compatible with Fig. 2.

Based on  the discussion above a grid of  force multiplier tables was computed for  10  blackbodies. Five blackbodies between 20 000 K - 50 000K were used for O-stars and five blackbodies between 100 000 K - 150 000 K were cut below 230 Å (fluxes set to zero) and used for WR-stars.  For an individual star, with a specific temperature, the force multiplier was interpolated from the grid.

\subsection{Dependence on particle density}
\begin{figure}
   \centering
   \includegraphics[width=9cm]{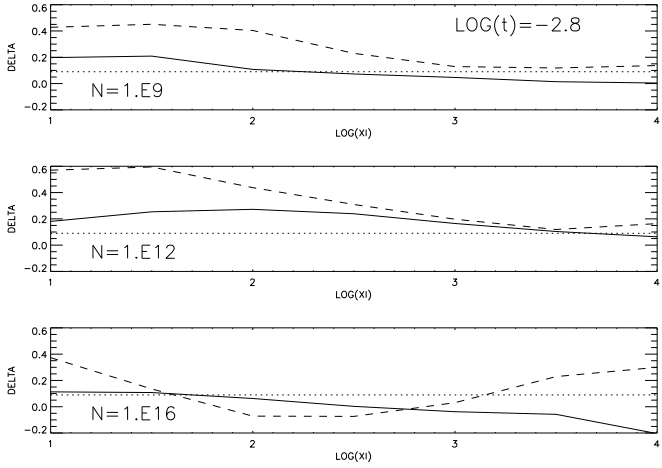}
     \caption{ The  $\delta$-parameter as a function of ionization parameter $\xi$ for  135 000 K cut blackbody (WR stars, solid) and 33000 K blackbody (O-stars, dash) at three $N$-values marked (cm$^{-3}$). The absorption parameter is fixed at log($t$) = -2.8. The dotted lines show $\delta$ = 0.11. }
         \label{delta}
   \end{figure}

 The particle number density $N$ is involved explicitely in $FM$ (resulting in a rather weak dependence)  and  in the ionization parameter $\xi$.  Fig.\ref{delta} shows  the $N$ - dependence  for a typical WN-star (135 000 K cut blackbody) and O-star (33 000 K blackbody) as a function of $N$ and $\xi$ at fixed log(t)=-2.8. The $\delta$-parameter  shown characterises the density-dependence of force multiplier and is defined as the derivative
\begin{equation}
\delta = d[log(FM)]/d[log(N)]
\end{equation}
Results are shown  for different values of $\xi$ and $N$ at log($t$)=-2.8.
  The value $\delta$=0.11, found  by Abbott (1982) using Mihalas and Kurucz atmospheres between effective temperatures 20 000 K - 60 000 K, is shown by horizontal dotted lines.  
 
\section{The Numerical Method}
\subsection{Wind Flow Equations}

Only the supersonic part of the wind flow was considered (above $\delta$(r/R$_{star}$) = 0.01). 
The two equations which determine the flow are  $\it{mass}$  $\it{conservation}$ 
\begin{equation}
\dot M = M_{\mathrm{{dot}}} = 4\pi  r^{2} \rho v  
\end{equation}
and  $\it{momentum}$  $\it{balance}$  (neglecting  gas pressure) 
\begin{equation}
 vdv/dr = GM/r^{2} [\Gamma_{e}(1 + g_{line}  ) - 1] \, {\rm}
\end{equation}
\begin{equation}
 g_{line} =   FM(r)F_{d}(r)F_{vel}^{2/3}  
\end{equation}
\begin{equation}
\Gamma_{e} = \sigma_{e}L/(4\pi GMc)  .
\end{equation}
t, $\xi$ and $N$ can be computed locally, hence, $FM$=$FM(r)$.
The momentum equation was derived by Sundqvist et al
 (2014) to account for  velocity-clumping in the
CAK-formalism.  To separate better from $F_{vol}$ we shall frequently use $FVEL$ instead of $F_{vel}$.

The force multiplier $FM$, computed in Ch.2,  is multiplied by two terms reducing the upwards force:
   finite  disc  correction  factor $F_{d}(r)$ and 
the  normalised velocity  filling  factor $F_{vel}$ (Venetian blind effect). 
According to Sundqvist et al. (2014) this  modified line force is implemented  also into the hydrodynamics code  VH-1 (developed by J.Blondin and collaborators).
$F_{vel}$  is defined as 
\begin{equation}
  F_{vel} = \delta v/(\delta v + \Delta v)  \,
\end{equation}
$\delta v$ is the velocity span  inside  a clump  and   $\Delta v$  is the  velocity separation  between  two individual clumps.  In this treatment statistical averages are used. 
The exponent 2/3  in $F_{vel}^{2/3}$ is the $\alpha$-parameter of CAK-formalism, the slope of line strength distribution function.    
The value of the finite  disc  correction  factor $F_{d}(r)$  is around 0.7 at stellar surface and rises to  1 at $r/R_{star}$ = 1.5 and remains constant thereafter. 
\subsection{Solution of Wind Equations}

The wind equations   were integrated numerically  using the IDL software,  resulting in velocity stratification v(r)  which was compared with the widely used model 
\begin{equation}
 v_{model} = v_{\mathrm{{inf}}}(1 - B/x)^\beta \,
\end{equation}
\begin{equation}
 B = 1. -  (v_{in}/v_{inf})^{1/\beta}   
\end{equation}  
\begin{equation}
 x =  r/R_{star} .
\end{equation}

The integrations were started above the subsonic region at $x$ = 1.01   and ended at  $x$  = 10,  using 1000 grid points with  $\delta x$ = 0.01. The subsonic part of the wind enters only via the boundary condition  $v_{in}$ = 10 km/s in eq 13. This will be discussed in Ch. 3.3 in context with the Potsdam-model WNE 16-20.
For each star  $M, R, L$ and chemical composition were used as input-values. The     $\beta$-parameter was fixed to 0.6. The effect of this choice will be discussed in Ch.6. 
The resulting $v(x)$ was then compared with the model $v_{model}(x)$ . The unknown free parameters mass loss rate $M$$_{dot}$ and $v_{inf}$  were  iterated using the IDL-procedure mpcurvefit.pro (written by Craig Markward), to get ($v(x)$ -  $v_{model}(x)$)/$v_{model}(x)$  to approach  zero at each $x$.   

The procedure mpcurvefit.pro  performs Levenberg-Marquard least squares fit, no weighting was used. 
The  $\chi^2$-value (chi2 in Tables A1 and A2) gives the goodness of fitting
 between the  computed velocity-stratification and  the velocity law ( DOF=998). 
The formal 1-sigma errors of each parameter were computed from the covariance matrix.
For this reason, these errors may not represent the true  parameter uncertainties.
For example, the small errors for computed $M$$_{dot}$-values in Table A2 may be underestimated.

The  effect of wind  clumping was introduced by multiplying the number density $N$ in the ionizing parameter $\xi$ by the factor  $1/F_{vol}$ where  $F_{vol}$  is  the clump $\it{volume \, filling \, factor}$.   The force multipliers were  originally computed using point-like radiation sources with the $\it{dilution \, factor}$ $r^{-2}$. However, in finite disk case in stellar atmospheres this dilution factor should be replaced by  $W(r)$  given by Mewe and Schrijver (1978) and used by Vink (2000): 
\begin{equation}
 W(r) = 0.5[1 - \sqrt(1 - (R_{star}/r)^2)] .
\end{equation}

Hence, the local  ionization parameter in the wind was computed from 
\begin{equation}
\xi (r) =  L_{ly}W F_{vol}/(N_{wind}R_{star}^{2}) .
\end{equation}

Here  $N_{wind}$ is the number density computed from mass conservation of the wind (eq.7).
This is a good assumption  if  the interclump space is empty and all mass is in the clumps. 

 \subsection{Comparison with the Potsdam WR-model WNE 16-20. Justification for the use of  cut black body.}
The wind equations were solved in the supersonic part of the wind (above  r/R$_{star}$=1.01) and assuming v$_{in}$=10 km/s as the boundary condition (eq.13).   Here these approximations are compared with the Potsdam model WNE 16-20  (Todt et al.  \citeyear{todt}, http://www.astro.physik-uni-potsdam.de/ ~wrh/POWR/WNE/16-20). The model parameters are: T=141 300 K, M = 12.02, R = 0.748, log(L) = 5.3, log(Mdot) = -4.83, v$_{inf}$ = 1600 km/s, $\beta$ = 1.0.  In the Potsdam NLTE modeling the  $\beta$-law was applied above the quasi-hydrostatic region. The Potsdam data base gives model stratifications and the outcoming spectrum. These are used here and compared with the methods of the present paper.

\begin{figure}
   \centering
   \includegraphics[width=9cm]{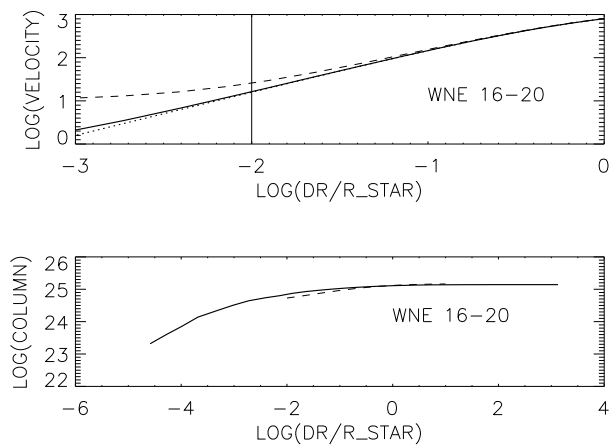}
  \caption{Upper plot: Logarithm of wind velocity (km/s) vs logarithm of height above stellar surface (r/R$_{star}$-1) 
for the Potsdam WR-model WNE 16-20 (solid line). The dotted line shows the Potsdam model above the subsonic region. The dashed line is the model from the present study. The vertical line marks the height above which the wind equations were applied. Lower plot: Logarithm of the accumulated column density vs logarithm of height for Potsdam model  WNE 16-20 (solid line). The dashed line shows the model from the present study.  } 
         \label{wnestrati}
   \end{figure}

The numerical method of Ch.3 explained above was applied to the WNE 16-20 parameter values keeping them constant without the iteration loop. The wind velocity stratification is shown in   Fig.\ref{wnestrati} (upper plot, dashed line). 
The lower plot shows the accumulated column density. The agreement above  r/R$_{star}$ =1.01 (the vertical line in the upper plot), where our flow equations were applied, is good.  The present paper  approximates the subsonic region by the boundary condition v$_{in}$ = 10 km/s and $\beta$ = 0.6.  Thermal and sound velocities are around this value. This enters into our treatment  via the column density estimate below  r/R$_{star}$ =1.01 (3E24  cm$^{-2}$) and the $\beta$-law. For the sample of Table A2 the rise of velocity is steeper, at  r/R$_{star}$=1.01  around 100 km/s. Hence, their treatment is well above the subsonic region. 

\begin{figure}
   \centering
   \includegraphics[width=9cm]{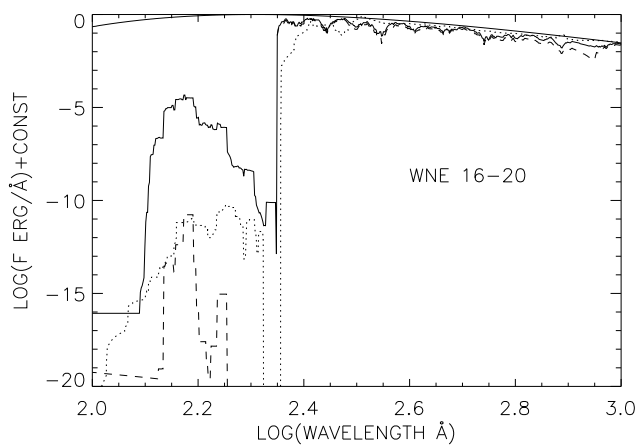}
  \caption{Outward fluxes for the  model WNE 16-20 smoothed by 10 Å boxcar.  The dotted line shows the Potsdam data base  emergent model (outside the wind). The present study gives the fluxes  at  r/R$_{star}$-1 = 0.01  (solid line) and at    r/R$_{star}$ = 10 (dashed line). The broad  smooth spectrum (thin solid line) is 141.3 kK blackbody used at  r/R$_{star}$ = 1 in the present study.                                                     }
         \label{wnespectra}
   \end{figure}

 Fig.\ref{wnespectra} shows the radiator fluxes smoothed by 10 Å boxcar for WNE 16-20 of the present treatment  at   r/R$_{star}$-1 = 0.01 ( heavy solid line) and  r/R$_{star}$-1 = 10 (dashed line) and compared with the Potsdam data base emergent flux (dotted line).  The 141.3 kK black body (the initial helium-star spectrum) is shown by the thin solid line.  The radiaton spectrum is  cut  below 230 Å  by 5 orders of magnitude already at the base of the supersonic region where the present treatment starts. Hence, the use of cut black bodies for computation of line force of WR-stars  is justified.
The removal of soft X-rays enhances the number of absorbers and consequently the line force. This is crucial for the wind acceleration in the supersonic region of Wolf Rayet stars..

\section{ Stellar data}
The O-star sample  consisted of 20 stars  from Howarth \& Prinja (1989, every 10th star from their list), 15 stars (models) from Kricka and Kubat (2017) and 5 low luminosity stars from Marcolino et al. (2009).  The Wolf Rayet sample consisted of 29 Nitrogen type (WN) stars from   Nugis \& Lamers (2000), 9 hottest (T$_{eff}$ over 80 000K)  WNE-stars from Hamann et al. (2006) and 16 hottest   WNE-stars (in LMC) from Hainich et al. (2014). 
The Wolf-Rayet companion of the X-ray binary Cygnus X-3  (WN-type) was added from  Vilhu et al. (2009). 
 
The data used  were the basic stellar parameters: mass $M$, radius $R$, luminosity $L$ and chemical composition.  These are given in Tables  A1 and A2 for the program stars with the computed values of   $M$$_{dot}$ and $v_{inf}$ (see Ch.5). The stars are shown in Fig. 7, a sort of HR-diagram. 

Since only basic parameters (M, R, L, composition) were used as inputs  for wind equations one might have used any grid of these parameters (e.g. from stellar evolutionary calculations). However, the above data were chosen because, besides the basic parameters, also observed and/or predicted mass loss rates and velocities  were given in quoted papers, permitting comparisons.

For O-stars solar abundances (Asplund et al. 2009) were used (by mass X =  0.71, Y = 0.27, Z = 0.015) while for Galactic WR/WN-stars  Hydrogen deficient abundances  were adopted as a mean value from  Nugis \& Lamers (2000)  (X = 0.085, Y = 0.9 and Z = 0.015 ).  
As found in Ch. 2.2  the use of  these abundances  result in rather similar force multipliers  but have some influence  on gas molecular weight and $\sigma$$_{e}$ . For O-stars  $\sigma$$_{e}$ = 0.33 while for WR-stars  $\sigma$$_{e}$ = 0.22. For  LMC-stars  smaller Z = 0.006 was adopted (Hainich et al. 2018) lowering the force multiplier respectively.

\begin{figure}
   \centering
   \includegraphics[width=9cm]{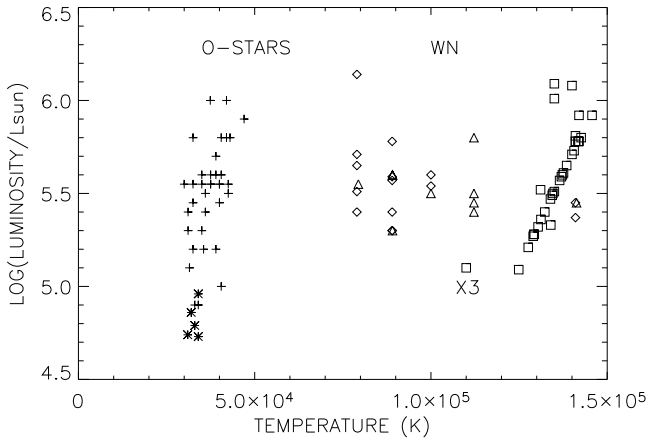}
     \caption{Effective temperature - luminosity diagram of the program stars. O- and WR-stars are well separated in temperature. Cyg X-3 is marked by 'X3'.Meaning of symbols (see Tables A1 and A2 reference numbers):  plus: 1 and 2; star: 3; square: 4 and 7; triangle: 5; diamond: 6 }
         \label{hr}
   \end{figure}


%
\section{Results. Computed mass loss rates.}

\subsection{O-stars and the weak wind problem}

For the O-stars (see Ch.4 and Table A1)  the wind equations    were solved  using Asplund (2009) solar abundances with  heavy metal mass fraction Z = 0.015. The input parameters were masses,  radii  and luminosities.  The $\beta$-exponent in eq. 12 was fixed to 0.6.  The solution was the mass loss rate  $M$$_{dot}$ and wind velocity at infinity $v$$_{inf}$.  The mass loss rates and wind velocities from the references of Table A1 were used  as  starting points for iterations.   $FVEL$ ($F$$_{vel}$)  in eq. 9 was set to 1.  Typically less than 10 iterations were needed to achieve  chisquares less than 1 (DOF=998).

\begin{figure}
   \centering
   \includegraphics[width=9cm]{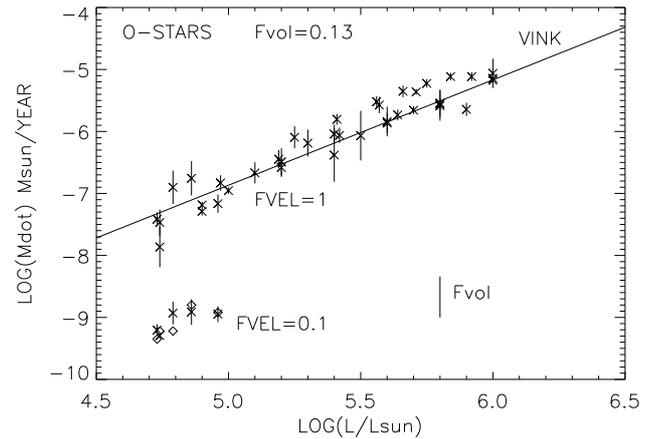}
     \caption{Mass loss rate  vs Luminosity for galactic O-stars (logarithmic scales, Table A1).  Solutions of wind equations for 20 stars from Howarth and Prinja (1989), 15 models from Krticka and Kubat (2017) and 5 low luminosity stars   from Marcolino et al. (2009) are shown with crosses with  error bars.  Asplund et al.  (2009) solar abundances were used  with Z=0.015. The clump volume filling factor $F$$_{vol}$ = 0.13. For the Marcolino's low luminosity  stars  solutions with the velocity filling factor $FVEL$=0.1 are included. The Vink-prediction line (Vink et.al. 2001) is added as the solid line. The diamonds show Marcolino's observations.  The vertical stick marked by Fvol shows the upwards effect when changing the volume filling factor from 1 to 0.1. }
         \label{ostars}
   \end{figure}

The results are shown in Fig. \ref{ostars} where the canonical Vink-prediction (Vink 2001) is shown by the solid line. This prediction  requires small volume filling factor $F$$_{vol}$ = 0.13. With unity  filling factor the results in   Fig.\ref{ostars} should be shifted down approximately by the amount of the stick shown in the figure (marked by Fvol).

In  late main sequence O-stars the observed mass loss rates are much lower than
predicted, contrary to giants and supergiants with higher luminosity. This is a manifestation of the 'weak wind problem'.  
Leakage of light associated with porosity in velocity space could lead to lowering of starlight power and consequently to lower mass loss rate.  This will be the case particularly for small clump spans (eq. 18).  
To test this idea, small values of $FVEL$ = 0.1 were applied to the main sequence low luminosity stars of Marcolino et al. (2009). With this value, 
the wind solutions matched well with the observations (see Fig.  \ref{ostars}) and can in principle solve the weak wind problem.  
This raises the question of the physical reason behind this porosity in velocity space:  why do main sequence dwarfs differ 
from giants and supergiants in this respect?  

For  low luminosity  stars  the mass loss rate depends on $FVEL$ as log($M$$_{dot}$) = const + 1.5$\times$log($FVEL$). 
For both our wind solutions and Vink-predictions the luminosity dependence of mass loss rate  is steep: log($M$$_{dot}$) = const + 1.7$\times$log($L$) . For Wolf Rayet stars the slope is  smaller  (1.2, see next Chapter).
Table A1 gives the results for O-stars with $F$$_{vol}$ = 0.13 and $FVEL$=1 except for main sequence Marcolino et al stars  for which  $FVEL$=0.1 was used.

\subsection{WR-stars and the momentum problem.}

The CAK-theory has been unable to explain the massive  wind of WR-stars and this is called a 'momentum  problem'. 
This name comes from the ratio of wind momentum to radiation momentum $\eta$ = $M$$_{dot}$v$_{inf}$/($L$/c) which is much larger than 1 for WR-stars. However, as pointed by Gayley et al. (1995), the problem may be  an opacity problem of simply identifying enough lines.    We  recomputed the force multipliers in Chapter 2  and found that, indeed, for realistic WR-spectra the line force is much stronger than predicted by CAK.  

 In Chapter 2 it was demonstrated that a cut blackbody suits well as the radiator of WR stars and was  used when solving the wind equations.  Fig.\ref{fm2} shows quantitatively how the force multiplier of  cut 135 000 K blackbody is over 10 times larger than that of a pure blackbody.   Figs \ref{elemcontri} and  \ref{wlcontri} show this as a function of element atomic number and wavelength, respectively. The soft X-ray suppression clearly vanishes when flux below 230 Ä is absorbed away.

\begin{figure}
   \centering
   \includegraphics[width=9cm]{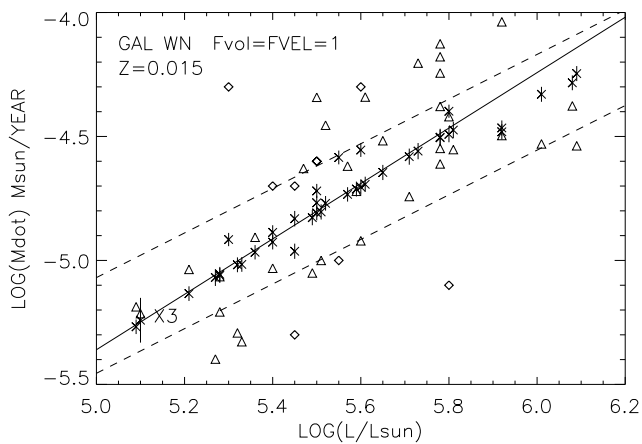}
     \caption{Mass loss rate vs Luminosity for galactic WN-stars (logarithmic scales, Table A2). Solutions of wind equations for 29 WN-stars from Nugis and Lamers (2000), Cyg X-3 from Vilhu et al. (2009),  and 9  hottest Hydrogen-deficient WN-stars from Hamann et al. (2006) are shown with crosses and with (small)  error bars  (using Z=0.015). Both volume and velocity filling factors were set to 1 ($F$$_{vol}$ = $FVEL$ = 1).  Triangles show the Nugis and Lamers (+ Cyg X-3) predictions and the solid  diagonal line their linear fit. The  diamonds show Hamann-predictions. The two dashed  lines  are Hainich-prescripions (Hainich et al (2014)) for two heavy element contents  Z = 0.006 (lower line) and 0.015 (upper line).  Cyg X-3 is marked by 'X3'.  }
         \label{galwne}
   \end{figure}

\begin{figure}
   \centering
   \includegraphics[width=9cm]{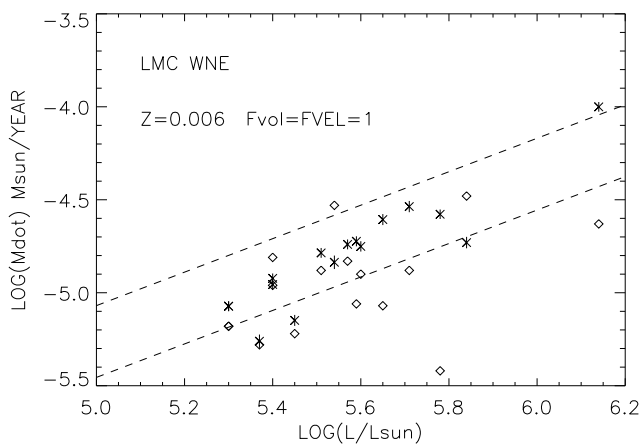}
     \caption{Mass loss rate vs Luminosity for Large Magellanic Cloud (LMC)  WNE-stars (logarithmic scale, Table A2). Solutions of wind equations for 16  Hydrogen deficient hottest stars from Hainich et al (2014) are shown with crosses and with (small) error bars for Z = 0.006. Both volume and velocity filling factors were set to 1 ($F$$_{vol}$ = $FVEL$ = 1). The dashed lines are the Hainich-prescriptions for two heavy element contents Z = 0.006 (lower) and 0.015 (upper). The diamonds show the observed values of Hainich et al. (2014).  }
         \label{lmcwne}
   \end{figure}

 As in the case of O-stars we need only the masses, radii, luminosities and chemical composition (plus the form of velocity-law, $\beta$-law with fixed $\beta$) to solve the wind equations and obtain mass loss rates and wind velocities. For Galactic targets Asplund et al. (2009) abundances with Z = 0.015 were used while for the LMC-stars Z = 0.006 was adopted (Hainich et al. 2015).  The results are shown in  Fig.\ref{galwne} (Milky Way stars) and  Fig.\ref{lmcwne} (LMC-stars). The results of all WN stars are given in Table A2 ($F$$_{vol}$ =  $FVEL$=1).  Around 10 iterations were needed to achieve chisquares less than 5 (DOF=998).

In  Figs \ref{galwne} and \ref{lmcwne} the Hainich-prescriptions (Hainich et al. 2014) are shown by dashed lines for two heavy element mass fractions (Z = 0.006 and 0.015). Mass loss predictions ('observations') are shown by triangles and diamonds. In  Fig.\ref{galwne} the solid line is a linear fit to the Nugis and Lamers data.  The  luminosity dependence is less steep than for O-stars: log($M$$_{dot}$) = const + 1.2$\times$log($L$) .   The Z-dependence is less steep than in the Hainich-prescription: log($M$$_{dot}$) = const + 0.42$\times$log(Z).

The predictions (observations) in Figures  \ref{galwne} and   \ref{lmcwne} show  very large scatter. A part of this  can be due to real  differences  in clumping parameters ($F$$_{vol}$ and $FVEL$) which may vary from star to star. This will be discussed in Ch. 6.

\subsection{Wind stratifications. Computed  wind velocities.
}
\begin{figure}
   \centering
   \includegraphics[width=9cm]{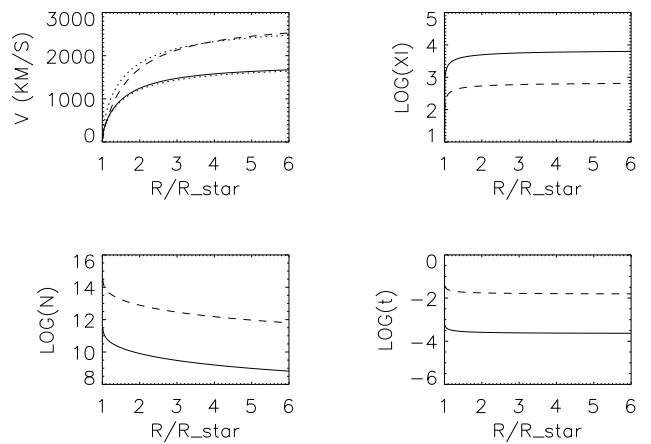}
  \caption{ Fitting results for  parameter-stratifications for means  of Howarth and Prinja (1989) O-stars in Table A1 (solid lines) and
               Nugis and Lamers (2000) WR-stars in Table A2 (dashed lines).  Upper left: wind velocity vs radial distance. The dotted lines (hardly visible) show the model. Upper right:  ionizing parameter log($\xi$) vs radial distance. Lower left: particle number density vs radial distance. Lower right: absorption parameter log($t$) vs radial distance.}
         \label{strati}
   \end{figure}

Examples of  parameter-stratifications  of our solutions  are shown in   Fig.\ref{strati} for means of  Howarth and Prinja (1989) O-stars and Nugis and Lamers (2000) WR-stars.  The differences between O- and WR-stars are due to  much larger mass loss rates of WR-stars which reflects in all the parameters shown.  Due to the simple modeling one can not expect extremely good fits. In the iterations chi squares (DOF = 998) for O-stars were  $\le{1}$ while for WR-stars chi squares $\le{5}$ were accepted. 

\begin{figure}
   \centering
   \includegraphics[width=9cm]{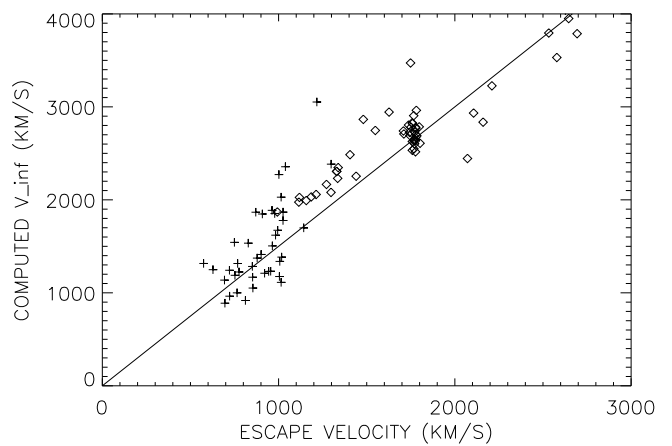}
     \caption{ Computed wind velocities at infinity versus escape velocities for the solutions of program stars (Tables  A1 and A2). Plusses: O-stars, open diamonds: WR-stars.  The solid line has a slope 1.5. }
         \label{vesc}
   \end{figure}

 The fitting results are collected in  Fig.\ref{vesc} showing $v$$_{inf}$ versus the escape velocity $v$$_{esc}$.  The correlation is rather tight  with
$v$$_{inf}$ = 1.5$v$$_{esc}$.  The escape velocity includes the reducing effect of Thomson scattering on gravity by 
\begin{equation}
 v_{\mathrm{{esc}}} = sqrt[2GM/R(1-\Gamma)]  
\end{equation}
 where   $\Gamma$  is computed from eq.10 (Kudritzki and Puls (2000), Nugis and Lamers (2000)).   

In principle  $v$$_{inf}$ can be changed, keeping $M$$_{dot}$ more or less  unchanged, by modifying the input parameters $\beta$, $F$$_{vol}$ and $F$$_{vel}$ = $FVEL$ (see next Chapter).  The computed  $v$$_{inf}$-values for O-stars (Table A1) match with the observed ones given in refs. 1-3 of Table A1.  The computed  $v$$_{inf}$-values for WN-stars (Table A2) are as an average 1.5 times larger than the 'observed' ones given in refs. 4-6 of Table A2 and which scatter around the escape velocities.   Using  smaller $\beta$ (instead of 0.6)  this difference becomes smaller,  $v$$_{inf}$ is more $\beta$-dependent than   $M$$_{dot}$ (see Table 1). 

\section{Discussion. Effect of parameters.}

XSTAR is designed to calculate the ionization and thermal balance in gases exposed to ionizing radiation. It contains a relatively complete collection of relevant atomic processes and corresponding atomic data. It does not, by itself, include processes associated with wind flows, such as shock heating, adiabatic cooling, or time-dependent effects. It also has a simple treatment of radiation transport which provides approximate results  in situations involving truly isotropic radiation fields or line blanketing, for example. 

When computing the ionization parameter $\xi$ for small  filling factors it was assumed that all the mass is in the clumps, without any interclump  low density gas with high ionization.  This simplified the method but underestimates the ionization if the interclump medium is  not empty. 

The wind equations  included three fixed parameters: $F$$_{vol}$, $FVEL$ and $\beta$ (see eqs 9, 12 and 16).  The value 0.13 of $F$$_{vol}$ for O-stars was selected to fit  mass loss solutions with Vink-prediction (Vink et al. 2001)  (in addition  $FVEL$ = 0.1 for weak wind stars, see Section 5.1). For WR-stars it was sufficient to select   $F$$_{vol}$ =  $FVEL$ = 1  to  match the computed mass loss rates with the Hainich-prescriptions (Hainich et al. 2014) and  with Nugis and Lamers (2001) predictions.  Computed  $v$$_{inf}$-values of O-stars are similar to the observed ones while for WN-stars the 'observed' values are by a factor 1.5 smaller than the computed ones in Table A2 (see Ch. 5.3). 

As an example, the computed mass loss rate of Cyg X-3 is very close to that found from orbital period change $P$/$P$$_{dot}$ = 850000 y (Kitamoto et al. 1995, Ergma and Yungelson 1998). The observed  $v$$_{inf}$ of Cyg X-3 is uncertain but Vilhu et al. (2009) give for the Si XIV emission line (arising in the photoionized wind) FWHM = 1850 $\pm{200}$ km/s and for the P Cygni absorption component at -900 km/s FWHM=750 $\pm{200}$ km/s.  These values point to large maximum velocity but  somewhat smaller than the computed one  given in Table A2  (2533 $\pm{205}$ km/s). The X-rays from the compact star may influence these values (to be studied later). If the X-ray suppression on the face-on side in Cyg X-3 is significant it may act in a similar way as lowering FVEL (see Table 1). The mass loss rate will be smaller but  this  may  be compensated by lower wind velocity because the accretion  is very sensitive on the wind velocity.

 To obtain  a quantitative estimate of the  influence of these parameters we computed the values by which   $M$$_{dot}$  and $v$$_ {inf}$ should be multiplied  if values other than the baseline ones ($\beta$ = 0.6,  $F$$_{vol}$  = $FVEL$ = 1)  are used. These are given in  Table 1. The values are  given  for the means of  O-star and WR-star samples.  Lowering of   $F$$_{vol}$ or $FVEL$ increases or decreases  $M$$_{dot}$, respectively. Increasing  of $\beta$  changes $M$$_{dot}$ and   $v$$_{inf}$ but in opposite way. In particular in WR-stars    $v$$_{inf}$  is sensitive on 
 $\beta$. 
\begin{table}
\begin{minipage}[t]{\columnwidth}
\caption{Effect of parameters  ($\beta$,  $F$$_{vol}$  and   $FVEL$ ) on  $M$$_{dot}$  and $v$$_ {inf}$ for the  means of Howarth and Prinja (1989, O-stars) and Nugis and Lamers (2000, WN-stars) samples. Table gives the values by which   $M$$_{dot}$  and $v$$_ {inf}$ should be multiplied  if  values other than the baseline values ($\beta$=0.6,   $F$$_{vol}$  = $FVEL$ =1) are used, as indicated in the Table.  
}
\label{lineint}
\centering
\renewcommand{\footnoterule}{}  
\begin{tabular}{lccccccc}
&&&$\beta$=1 &$F$$_{vol}$=0.1  & $FVEL$=0.1 & \\
\\
&O-stars& $M$$_{dot}$&0.66&4.32&0.02&\\
& &$v$$_ {inf}$&1.42&1.26&1.35&\\
&WN-stars&$M$$_{dot}$ &0.75&1.86&0.08&\\
&&$v$$_ {inf}$&2.70&1.05&0.89&          \\
\\
\end{tabular}
\end{minipage}
\end{table}

The velocity clumping parameter $FVEL$ is related to velocity-gradients.
 Numerical simulations show that after a local velocity jump there follows a region with  smaller velocity gradient producing density clumps (Sundqvist et al. 2010).   In the smooth wind case (with monotonic velocity law)   $\delta$v$_{sm}$/$\Delta$v =  clump size /clump separation =   $\delta$r$_{sm}$/$\Delta$r.   Assuming spherical clumps the volume filling factor is by definition $F_{vol}$ = 4/3$\pi$($\delta$r/2)$^3$/$\Delta$r$^3$. Using this,  $F_{vel}$   can be rewritten from  eq. 11 as
\begin{equation}
 F_{\mathrm{{vel}}} = FVEL = span \times 1.24 F_{vol}^{1/3}/(1 + span \times 1.24 F_{vol}^{1/3})  
\end{equation}

 Here $span$ =  $\delta$v/$\delta$v$_{sm}$ = velocity span in clumps relative to the smooth wind case. Hence,  for the best fit with Vink-prediction of O-stars ( $F$$_{vol}$ = 0.13),  $FVEL$ = 0.38 and 0.98 if the span is 1 and 100, respectively. This may be one area when discussing why low luminosity weak wind stars could have small $FVEL$-values: sudden change in the wind vorosity when moving from O-type giants towards main sequence.


\section{Conclusions}

Past work has demonstrated that the CAK-method is succesful for
 giants and supergiants of normal OB-stars but failed to explain the weak winds of main 
sequence low luminosity OB-stars (‘weak wind problem’). Further, CAK was never 
applied seriously to WR-stars due to the 
 ‘momentum problem’.  In the present paper the CAK-method has been tested in this context by
recalculating  force multipliers and integrating the 
wind equations for a sample of O- and WR stars.  Mass loss rates and wind velocities comparable with the observed ones were obtained and as a byproduct solutions for the momentum and weak wind problems suggested.  

The work reported in this paper represents a comprehensive effort to model the winds of a large sample of hot stars. It also includes the effects of ionization on the force multiplier computed locally in the wind.
 Starting from the basic stellar parameters (mass, radius, luminosity, chemical composition)  the wind equations were solved  in the supersonic part and fitted with the velocity $\beta$-law with $\beta$=0.6  (eq. 12) and $v$$_{in}$ = 10 km/s, giving mass loss rates  ($M$$_{dot}$) and wind velocities  ($v$$_{inf}$) as the results.  These equations included both density and velocity clumpings ($F$$_ {vol}$ and $FVEL$=$F$$_ {vel}$).   The $\beta$-law used has no physical justification for WR-stars, neither fixed $\beta$= 0.6. However, this was used to show that the wind-acceleration works.  Besides the basic stellar parameters the mass loss rate depends also on the velocity law and the subsonic part boundary condition. These were approximated  by the  values of   $\beta$ and   $v$$_{in}$.

The results are given in Tables  A1 and A2 and Figures 8-12. The resulting mass loss rates were  compared with observations/predictions. O-stars require moderate density  clumping ($F$$_ {vol}$ = 0.13, $FVEL$ = 1) when  compared with Vink et al. (2001) predictions.  WR-winds can be modelled  with  $F$$_ {vol}$ = $FVEL$ = 1.  The results depend somewhat on the velocity model used (value of $\beta$ in eq. 9),  for $M$$_{dot}$  the   $\beta$-dependence is weaker  than  for $v$$_{inf}$ (see Table 3). 

The line force multiplier (radiative force) was computed with the XSTAR code and  atomic data base  as a function of local   parameters $t$  (Sobolev line absorption), $\xi$ (ionization parameter) and $N$ (particle number density). 
 The ionizing source (the stellar spectrum) was flexibly specified ranging from  blackbodies to realistic computed spectra. It was demonstrated that blackbodies can be used as radiators for O-stars while for WR-stars one should use cut blackbodies (fluxes cut to zero below 230 Å, the He$^+$ ionization limit).  
  
The flux below 230 Å  is crucial for the size of  force multiplier.  That part of radiation flux  (soft X-rays) can effectively suppress the radiative force. The lack of it in WR-stars, due to the strong absorption in this spectral region, increases the number of lines contributing to the line force  and avoids the X-ray suppression (see Figs 2-5). This makes possible to accelerate their massive winds and solves the momentum  and  single  scattering   limit  problems.  Hence, the 'momentum problem' is an opacity problem as suggested by Gayley et al. (1995).
      
A possible solution for the 'weak  wind problem' of low luminosity late O-stars was quantitatively studied. A small velocity filling factor  $FVEL$ = 0.1 solves the  problem but the physical reason behind this remains to be clarified.
As discussed in the previous Chapter the  $clumpspan$ (velocity span inside a clump relative to the smooth wind case) may be different in giants and main sequence O-stars.

\begin{acknowledgements}
  
\end{acknowledgements}

\begin{appendix}
\section{Targets}
\begin{table}
\begin{minipage}[t]{\columnwidth}
\caption{Basic parameters for O-stars with the results  of numerical iterations.    
 Mass $M$, radius $R$ and  luminosity $L$  are in solar units. The computed mass loss rate (Mdot)  is
in solar masses/year and velocity at infinity (Vinf) in km/s.  The errors (+-) are formal 1-sigma errors.
 Solar abundances were used. $F$$_{vol}$ = 0.13 and $FVEL$=1 except for main sequence stars of ref 3  $FVEL$=0.1.
For ref 1 and 3 the NAME is HD- or BD-number or the name of the target, or ref 3 it is the model number. chi2 is the chisquare of the fitting (between velocity and model velocity), DOF=998. }
\renewcommand{\footnoterule}{}  
\begin{tabular}{lcccccccc}

&NAME&$M$  &$R$  & Log($L$)  &  Log(Mdot) & Vinf&chi2&ref\\
&   108&61.0&16.1& 5.80 &-5.55+- 0.22 & 1867+- 886&0.6& 1\\
& 15137&29.0&15.4& 5.30 &-6.20+- 0.22 & 1315+- 185&0.2& 1\\
& 34656&32.0&10.5& 5.20 &-6.50+- 0.05 & 1175+- 160&2.5& 1\\
& 37468&25.0& 8.6& 4.90 &-7.30+- 0.07 & 1337+- 372&1.5& 1\\
& 46202&24.0& 8.1& 4.90 &-7.19+- 0.05 & 1112+- 162&2.5& 1\\
& 54662&45.0&12.8& 5.60 &-5.87+- 0.16 & 2028+- 646&0.5& 1\\
& 66811&74.0&18.9& 6.00 &-5.18+- 0.04 & 1620+-1005&0.8& 1\\
& 90273&31.0& 6.4& 5.00 &-6.96+- 0.05 & 2383+- 853&2.2& 1\\
& 93843&57.0&14.3& 5.80 &-5.57+- 0.09 & 2355+-1107&1.2& 1\\
&101131&56.0&15.0& 5.80 &-5.59+- 0.24 & 2271+-1004&0.4& 1\\
&11278.&26.0&11.9& 5.10 &-6.68+- 0.17 & 1168+- 162&0.4& 1\\
&148937&52.0&15.5& 5.70 &-5.66+- 0.06 & 1852+- 489&0.8& 1\\
&152247&32.0&17.3& 5.40 &-6.39+- 0.43 & 1542+- 704&3.2& 1\\
&153919&69.0&23.8& 6.00 &-5.07+- 0.24 & 1534+- 789&0.3& 1\\
&164492&34.0&12.9& 5.40 &-6.05+- 0.15 & 1412+- 330&0.7& 1\\
&168076&73.0&13.4& 5.90 &-5.65+- 0.11 & 3052+- 795&7.3& 1\\
&190864&44.0&13.8& 5.60 &-5.85+- 0.24 & 1887+- 682&0.4& 1\\
&203064&37.0&14.5& 5.50 &-6.08+- 0.40 & 1867+- 861&0.4& 1\\
&218195&28.0&12.6& 5.20 &-6.50+- 0.22 & 1283+- 243&0.4& 1\\
&592603&34.0& 8.7& 5.20 &-6.59+- 0.15 & 1698+- 576&1.2& 1\\
&  3255&16.4& 7.4& 4.74 &-7.48+- 0.21 & 1373+- 574&0.5& 2\\
&  3505&20.9& 8.3& 4.97 &-6.84+- 0.12 & 1210+- 266&1.1& 2\\
&  3755&26.8& 9.3& 5.19 &-6.46+- 0.15 & 1504+- 407&0.8& 2\\
&  4005&34.6&10.7& 5.42 &-6.07+- 0.12 & 1672+- 514&0.8& 2\\
&  4255&45.0&12.2& 5.64 &-5.74+- 0.09 & 1778+- 616&1.2& 2\\
&  3253&22.8&13.3& 5.25 &-6.10+- 0.18 & 1242+- 210&0.5& 2\\
&  3503&27.2&13.8& 5.41 &-5.81+- 0.09 & 1188+- 296&1.0& 2\\
&  3753&32.5&14.4& 5.57 &-5.58+- 0.13 & 1224+- 494&0.6& 2\\
&  4003&39.2&14.9& 5.71 &-5.37+- 0.05 &  917+- 420&0.6& 2\\
&  3001&28.8&22.3& 5.56 &-5.53+- 0.09 & 1315+- 259&1.3& 2\\
&  3251&34.0&21.3& 5.66 &-5.36+- 0.09 & 1248+- 288&1.2& 2\\
&  3501&40.4&20.4& 5.75 &-5.23+- 0.07 & 1136+- 363&0.8& 2\\
&  3751&48.3&19.7& 5.84 &-5.12+- 0.06 &  999+- 275&0.6& 2\\
&  4001&58.1&19.0& 5.92 &-5.12+- 0.07 & 1051+- 486&0.5& 2\\
&  4251&70.3&18.4& 6.00 &-5.15+- 0.07 & 1231+- 690&1.7& 2\\
&216898&17.0& 6.7& 4.73 &-9.20+- 0.09 & 1229+- 542&1.0& 3\\
&326329&19.0& 8.1& 4.74 &-9.29+- 0.04 & 1847+- 596&2.1& 3\\
& 66788&26.0& 8.7& 4.96 &-8.95+- 0.13 & 1384+- 716&0.8& 3\\
&ZetaOph&13.0& 8.8& 4.86 &-8.91+- 0.21 &  888+- 368&0.5& 3\\
&216532&12.0& 7.6& 4.79 &-8.93+- 0.18 &  964+- 329&0.5& 3\\

\end{tabular}
\tablebib{(1) ~\citet{howarth}; (2) ~\citet{keticka}; (3)~\citet{marcolino}};

\end{minipage}
\label{dataO}
\end{table}

\begin{table*}
\caption{Basic parameters for WN-stars  with the results of numerical iterations .     Mass $M$, radius $R$ and  luminosity $L$  are in solar units.  The computed mass loss rate (Mdot) is in solar masses/year and velocity at infinity (Vinf) in km/s.   The errors (+-) are formal 1-sigma errors. Hydrogen deficient abundances  were  used and for LMC WN-stars (ref. 6)  one third of the Galactic heavy element content was  adopted. $F$$_{vol}$ = $FVEL$=1.  chi2 is the chisquare of the fitting (between velocity and model velocity), DOF=998. For ref 4 the  NAME is the WR-number in the Sixth catalogue of WR-stars (van der Hucht et al. 1981), for ref 5 the WR-number in the Seventh catalogue of WR.stars (van de Hucht 2001) and for ref 6 the BAT99-number (Breysacher et al. 1999). }

\renewcommand{\footnoterule}{}  
\begin{tabular}{llllllllllccccccccc}
&NAME&$M$  &$R$  & Log($L$)  &  Log(Mdot) & Vinf&chi2&ref\\
&     2&10.0& 0.9& 5.27 &-5.07+- 0.03 & 3472+- 903&3.2& 4\\
&   127&10.8& 0.9& 5.33 &-5.02+- 0.03 & 2785+- 423&4.9& 4\\
&     1&15.2& 1.1& 5.57 &-4.73+- 0.03 & 2962+- 551&4.3& 4\\
&     6&15.6& 1.1& 5.59 &-4.71+- 0.03 & 2781+- 418&5.0& 4\\
&    31&10.1& 0.9& 5.28 &-5.06+- 0.03 & 2692+- 393&5.0& 4\\
&    51&10.2& 0.9& 5.28 &-5.06+- 0.03 & 2627+- 335&5.4& 4\\
&   151&18.5& 1.2& 5.71 &-4.58+- 0.03 & 2651+- 358&5.5& 4\\
&    10&10.8& 0.9& 5.32 &-5.02+- 0.03 & 2643+- 335&5.4& 4\\
&    21&12.0& 1.0& 5.40 &-4.93+- 0.03 & 2729+- 389&5.0& 4\\
&    97&11.3& 0.9& 5.36 &-4.97+- 0.03 & 2819+- 475&4.6& 4\\
&   133& 8.0& 0.7& 5.09 &-5.27+- 0.03 & 2725+- 416&5.1& 4\\
&   138&13.9& 1.0& 5.51 &-4.80+- 0.03 & 2603+- 300&5.7& 4\\
&   139& 9.3& 0.8& 5.21 &-5.13+- 0.03 & 2780+- 469&4.8& 4\\
&   141&15.8& 1.1& 5.60 &-4.70+- 0.03 & 2635+- 322&5.7& 4\\
&   157&13.5& 1.0& 5.49 &-4.83+- 0.03 & 2549+- 265&5.9& 4\\
&    24&48.0& 1.8& 6.01 &-4.33+- 0.03 & 3794+- 456&5.3& 4\\
&    25&57.0& 2.0& 6.09 &-4.25+- 0.03 & 4040+- 540&5.0& 4\\
&    47&40.0& 1.4& 5.92 &-4.48+- 0.03 & 3950+- 523&4.9& 4\\
&    67&16.1& 1.1& 5.61 &-4.69+- 0.03 & 2688+- 352&5.5& 4\\
&   115&13.6& 1.0& 5.50 &-4.81+- 0.03 & 2828+- 480&4.6& 4\\
&   136&19.1& 1.2& 5.73 &-4.56+- 0.03 & 2700+- 391&5.3& 4\\
&   153&14.0& 1.1& 5.52 &-4.77+- 0.03 & 2741+- 449&4.7& 4\\
&   155&17.0& 1.2& 5.65 &-4.65+- 0.03 & 2770+- 408&5.1& 4\\
&    22&55.3& 1.9& 6.08 &-4.28+- 0.03 & 3787+- 293&5.9& 4\\
&    78&21.5& 1.3& 5.80 &-4.49+- 0.03 & 2608+- 300&5.8& 4\\
&    87&40.0& 1.5& 5.92 &-4.47+- 0.03 & 3530+- 249&6.2& 4\\
&    40&20.6& 1.3& 5.78 &-4.50+- 0.03 & 2635+- 355&5.5& 4\\
&   147&20.6& 1.3& 5.78 &-4.51+- 0.03 & 2668+- 371&5.4& 4\\
&   105&21.8& 1.3& 5.81 &-4.47+- 0.03 & 2516+- 254&6.1& 4\\
& Cyg X-3&10.0& 1.0& 5.10 &-5.24+- 0.03 & 2533+- 205&6.0& 7\\
&     1&15.0& 1.3& 5.40 &-4.89+- 0.03 & 2905+- 500&4.6& 5\\
&     2&16.0& 0.9& 5.45 &-4.96+- 0.03 & 3226+- 389&5.5& 5\\
&     6&19.0& 2.6& 5.60 &-4.55+- 0.03 & 2347+- 464&3.7& 5\\
&     7&16.0& 1.4& 5.45 &-4.83+- 0.03 & 2716+- 411&5.0& 5\\
&    18&17.0& 1.5& 5.50 &-4.77+- 0.03 & 2804+- 449&4.6& 5\\
&    37&17.0& 1.9& 5.50 &-4.72+- 0.03 & 2746+- 592&3.6& 5\\
&    46&25.0& 2.1& 5.80 &-4.40+- 0.03 & 2943+- 660&3.7& 5\\
&    36&13.0& 1.9& 5.30 &-4.92+- 0.03 & 2485+- 537&3.6& 5\\
&    44&18.0& 3.1& 5.55 &-4.58+- 0.03 & 2057+- 376&3.7& 5\\
&     1&12.0& 1.9& 5.30 &-5.07+- 0.03 & 2306+- 471&4.0& 6\\
&     2&13.0& 0.8& 5.37 &-5.26+- 0.04 & 2444+- 279&7.7& 6\\
&     3&16.0& 3.0& 5.51 &-4.79+- 0.03 & 1992+- 441&3.2& 6\\
&     5&15.0& 0.9& 5.45 &-5.15+- 0.03 & 2933+- 260&5.2& 6\\
&     7&25.0& 1.1& 5.84 &-4.73+- 0.03 & 2835+- 391&5.3& 6\\
&    15&17.0& 2.6& 5.57 &-4.74+- 0.03 & 2165+- 419&3.6& 6\\
&    19&39.0& 6.3& 6.14 &-4.00+- 0.03 & 1869+- 524&2.9& 6\\
&    24&17.0& 2.0& 5.54 &-4.84+- 0.04 & 2865+- 701&3.2& 6\\
&    36&21.0& 3.8& 5.71 &-4.54+- 0.03 & 2026+- 442&3.0& 6\\
&    37&19.0& 3.6& 5.65 &-4.61+- 0.03 & 1975+- 461&3.0& 6\\
&    41&18.0& 2.1& 5.60 &-4.75+- 0.03 & 2254+- 281&4.5& 6\\
&    47&18.0& 2.6& 5.59 &-4.72+- 0.03 & 2080+- 320&4.1& 6\\
&    48&14.0& 2.1& 5.40 &-4.96+- 0.03 & 2231+- 489&4.0& 6\\
&    51&12.0& 1.9& 5.30 &-5.07+- 0.03 & 2306+- 471&4.0& 6\\
&    57&14.0& 2.7& 5.40 &-4.92+- 0.03 & 2031+- 492&3.4& 6\\
&    66&35.0& 3.3& 5.78 &-4.58+- 0.03 & 2708+- 399&4.8& 6\\

\end{tabular}
\tablebib{
(4) ~\citet{nugis}; (5)~ \citet{hamann}; (6)~ \citet{hainich}; (7)~Cyg X-3~ \citet{vilhu}}
\label{dataWN}
\end{table*}
\end{appendix}
 \end{document}